\documentclass[aps,pre,twocolumn,showpacs,superscriptaddress]{revtex4-1}
\usepackage{graphicx}
\usepackage[colorlinks=true,linkcolor=blue,%
linkcolor=blue,pagecolor=blue,citecolor=blue,%
urlcolor=blue,anchorcolor=black,bookmarksnumbered=true,bookmarksopen=true]{hyperref}%
\renewcommand{\frac}[2]{\displaystyle{#1 \over #2}}
\begin{document}
\title{ONSET OF CAVITY DEFORMATION
UPON\\ SUBSONIC MOTION OF A
PROJECTILE IN A FLUID\\ COMPLEX
PLASMA}
\author{D.~I.~Zhukhovitskii} \email{dmr@ihed.ras.ru}
\affiliation{Joint Institute of High Temperatures, Russian
Academy of Sciences, Izhorskaya 13, Bd.~2, 125412
Moscow, Russia}
\author{A.~V.~Ivlev}
\affiliation{Max-Planck-Institut f\"{u}r extraterrestrische
Physik, Giessenbachstrasse, 85748 Garching, Germany}
\author{V.~E.~Fortov}
\affiliation{Joint Institute of High Temperatures, Russian
Academy of Sciences, Izhorskaya 13, Bd.~2, 125412
Moscow, Russia}
\author{G.~E.~Morfill}
\affiliation{Max-Planck-Institut f\"{u}r extraterrestrische
Physik, Giessenbachstrasse, 85748 Garching, Germany}
\date{\today}
\begin{abstract}
We study the deformation of a cavity around a large projectile moving with subsonic velocity in the cloud of small dust
particles. To solve this problem, we employ the Navier--Stokes equation for a compressible fluid with due regard for friction
between dust particles and atoms of neutral gas. The solutions shows that due to friction, the pressure of a dust cloud at the surface of a cavity around the projectile can become negative, which entails the emergence of a considerable asymmetry of the cavity, i.e., the cavity deformation. Corresponding threshold velocity is calculated, which is found to decrease
with increasing cavity size. Measurement of such velocity makes it possible to estimate the static pressure inside the dust
cloud.
\end{abstract}
\maketitle

\section{\label{s1} INTRODUCTION}

Complex (dusty) plasma is a low-temperature plasma including microparticles. Due to electron absorption, particles acquire a
considerable electric charge. Thus, a strongly coupled Coulomb system is formed \cite{1,2,3,4,5,6,7,8,9}. Such plasma
represents a natural system which makes it possible to observe various collective phenomena at the level of individual
particles. In experimental setups, complex plasmas are usually studied in gas discharges at low pressures, e.g., in radio
frequency (RF) discharges. A large homogeneous bulk of complex plasma, which almost fills the entire discharge volume, can
be observed under microgravity conditions either in parabolic flights \cite{10,11,12,13,14} or onboard the International
Space Station (ISS) \cite{10,15,16,17,18,19}.

In recent studies, attention was focused on tracer particles or projectiles moving through a cloud of complex plasma. Such
projectiles are generated using controlled mechanisms of acceleration \cite{11,20}, or they can appear sporadically
\cite{19,21}. Projectiles moving with supersonic velocity lead to the formation of extended Mach cones; subsonic (slow) ones
produce localized disturbances of surrounding particles. In Ref.~\onlinecite{22}, it was suggested that the latter regime,
realized when a relatively large subsonic projectile moves in a dense cloud of smaller particles, can be well approximated
by a flow of an incompressible fluid. In the framework of the same hydrodynamic approximation, it was demonstrated that,
along with the regular neutral gas drag, there is an additional force exerted on the projectile due to friction between
neutral atoms and the particle fluid \cite{25}.

In this study, we try to estimate the deformation threshold for an (initially spherical) cavity around a subsonic projectile.
This deformation emerges abruptly as the projectile velocity increases, as it is seen in snapshots given in \cite{22,11}. To
solve this problem, we employ the Navier--Stokes equation for a compressible fluid taking into account friction between dust
particles and atoms of neutral gas. We do not treat the deformation self-consistently. Instead, we imply that the projectile
velocity is below the threshold value, so that a regular flow around a spherical cavity with no stall can be treated. The
solutions shows that due to friction, the pressure of a dust cloud at the boundary of the cavity behind the projectile can
become negative, which entails the formation of a microscopic void free from dust particles, i.e., the cavity deformation. This
occurs at some threshold velocity which decreases with increasing cavity size. Measurement of such velocity would make it
possible to estimate the static pressure inside the dust cloud.

The paper is organized as follows. In Sec.~\ref{s2}, we solve the non-stationary Navier--Stokes equation for an
incompressible particle fluid. In Sec.~\ref{s3}, the gas dynamics problem is solved and the corrections for fluid
compressibility to the velocity and pressure fields are calculated. Calculation results are compared with available
experimental data in Sec.~\ref{s4} and the results of this study are summarized in Sec.~\ref{s5}.

\section{\label{s2} INCOMPRESSIBLE FLUID
APPROXIMATION}

Consider an irrotational flow of incompressible particle fluid formed by the dust crystal melted around a projectile moving
with the velocity ${\bf{u}}(t)$ relative to the dust, where $t$ is the time.
The parameter of interaction between dust particles is large due to their low kinetic energy at room temperature. The corresponding parameter for the interaction between the projectile and dust particles can be defined as the ratio of characteristic Coulomb energy to the kinetic energy of a dust particle,
$\beta _{dp} = 2Z_p Z_d e^2 /\lambda M_d u^2$ (where $Z_p$ and $Z_d$ are the charges of the projectile and dust particle, respectively, in units of the electron charge $e$, $\lambda \simeq 6 \times 10^{ - 3} \;{\mbox{cm}}$
is the plasma (ion) screening length, and $M_d$ is the dust particle mass). Under typical conditions, this parameter is great, $\beta _{dp} \sim 100$ \cite{22}. In such a strongly coupled system, each particle (including the projectile) finds itself in the center of the spherically symmetric Wigner--Seitz cell. Since each cell is electrically neutral, the interaction of particles can be effectively reduced to that of cell surfaces, which can be modeled by hard spheres. This applies to the surface of a cell around the projectile as well. This approach is valid until the cavity deformation occurs; however, we will confine ourselves to the treatment of spherically symmetric cavities below the deformation threshold. We denote the radius of a cavity around the projectile by $R$. 

The Navier--Stokes equation describing the
velocity field ${\bf{v}}({\bf{r}},\,t)$ in the reference frame of the projectile has the form
\begin{equation}
\frac{{\partial {\bf{v}}}}{{\partial t}} +
({\bf{v}}\cdot\nabla ){\bf{v}} + \nu ({\bf{v}} + {\bf{u}})
= - \frac{{\nabla p}}{\rho }, \label{e1}
\end{equation}
where $\rho = M_d n_d$ is the mass density of dust fluid (assumed in this section to be constant), with $M_d$ and $n_d$
being, respectively, the mass and number density of dust particles of the radius $a_d$. Furthermore, $p(t,\,{\bf{r}})$ is
the dust pressure field, $\nu = (8\sqrt {2\pi } /3)\delta m_n n_n v_{Tn} a_d^2 /M_d$ is the friction coefficient with
$\delta \simeq 1.4$ being the accommodation coefficient \cite{6}, and $m_n$, $n_n$, $T_n$, and $v_{Tn} = (T_n /m_n )^{1/2}
$ are the mass, number density, temperature, and thermal velocity of neutral gas molecules, respectively. Equation
(\ref{e1}) assumes also that far from the projectile, dust particles are quiescent relative to neutral gas. It was shown in
\cite{22} that the approximation of nonviscous flow results in a fairly good description of trajectories of individual dust
particles; the estimate of the viscosity term for the dust particle fluid is indicative of the fact that in most cases, it is
small \cite{25}. This allowed us to omit it in Eq.~(\ref{e1}) and to confine ourselves to the nonviscous approximation.

For an incompressible fluid, the continuity equation is
reduced to
\begin{equation}
\nabla \cdot {\bf{v}} = 0. \label{e2}
\end{equation}
The boundary conditions for (\ref{e1}) and (\ref{e2}) are $\left. {({\bf{v}} \cdot {\bf{n}})} \right|_{r = R} = 0$, where
${\bf{n}} = {\bf{r}}/r$, and ${\bf{v}} = - {\bf{u}}$ at $r = \infty $. For an irrotational flow ($\nabla \times {\bf{v}} =
0$), we substitute ${\bf{v}} = \nabla \varphi - {\bf{u}}$ in Eq.~(\ref{e2}) to obtain the equation
\begin{equation}
\nabla^2\varphi=0,\quad \varphi(t,\,\infty)=0,\quad \left.\frac{\partial\varphi}{\partial{\bf n}}\right|_{r=R}=0, \label{e3}
\end{equation}
which has the solution \cite{24}
\begin{equation}
\varphi({\bf r }) = - \frac{{R^3 }}{{2r^2 }}{\bf{u}} \cdot{\bf{n}} \label{e4}
\end{equation}
or
\begin{equation}
{\bf{v}}({\bf r }) = -\frac{{R^3 }}{{2r^3 }}\left[{\bf u}-3{\bf n}({\bf n}\cdot{\bf u}) \right] -{\bf u}. \label{e5}
\end{equation}

The pressure field $p({\bf r})$ can then be found by substituting velocity (\ref{e5}) in Eq.~(\ref{e1}). This yields the
following pressure distribution at the spherical surface of a cavity $(r=R)$,
\begin{equation}
p = p_{\rm st} + \frac{{\rho \nu R}}{2}{\bf{u}} \cdot {\bf{n}}
+ \frac{{\rho u^2 }}{8}\left[ {9\frac{{({\bf{u}} \cdot
{\bf{n}})^2 }}{{u^2 }} - 5} \right] + \frac{{\rho
R}}{2}\dot{\bf u}\cdot{\bf n}, \label{e8}
\end{equation}
where $p_{\rm st}={\rm const}>0$ is the static pressure of dust at $r = \infty $.

\begin{figure}
\unitlength=0.24pt
\begin{picture}(500,700)
\put(-90,-40){\includegraphics[width=2.5in]{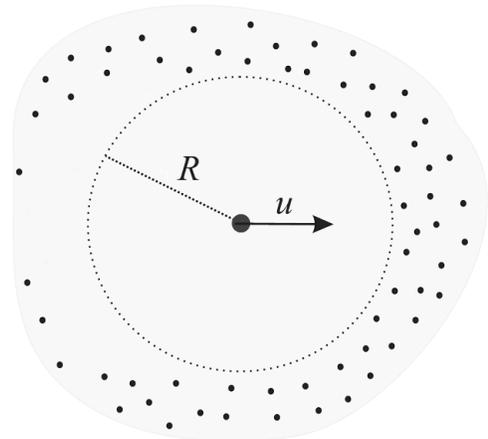}}
\end{picture}
\caption{\label{f1}Deformation of a cavity behind moving 
projectile (large bullet in the center) moving with the 
velocity $u$
 through the dust crystal (small bullets around) in a carrier 
gas. The projectile is surrounded by deformed cavity with 
initial radius $R$.}
\end{figure}

Consider the deformation of a cavity around a projectile propagating in the dust crystal (Fig.~\ref{f1}).
Here, we will neglect
finite fluid compressibility -- corresponding corrections are discussed in Sec.~\ref{s3}. Deformation can be caused solely
by the formation of a void in the stream of dust fluid because the interaction parameter $\beta _{dp}$ is large. Since no
evidence for an appreciable attraction between dust particles has been reported in the literature, we model the cloud of
dust particles by a system of hard spheres. In such a system, a cavity can be deformed by void formation if the
condition $p \le 0$ is satisfied over some area around a cavity. This condition corresponds to  cavitation in a
metastable fluid and to stall in gas dynamics.

Let us first consider the case $\dot u=0$. According to Eq.~(\ref{e8}), friction between dust particles and neutral gas
gives rise to a pressure increase in front of the projectile and a decrease behind it, so that $p({\bf r})$ reaches a
maximum at the ``front pole'' (where ${\bf u } \cdot {\bf n} = u$). By introducing the cosine of the polar angle,
$\cos\theta= {\bf u} \cdot{\bf n}/u\equiv\xi$, we readily derive
\begin{equation}
\frac{\partial p}{\partial \xi}=\frac94\rho u^2(\xi-\xi_{\rm cr}),
\end{equation}
where $\xi_{\rm cr}=-2\nu R/9u$. If $|\xi_{\rm cr}|>1$ then the pressure reaches a minimum at the ``rear pole''
($\xi=-1$), where $p_{\rm min}=p_{\rm st} - \rho u^2(9|\xi_{\rm cr}|/2 - 1)/2$. This corresponds to the low-velocity
regime $u < 2\nu R/9$, when the minimum pressure decreases (approximately) linearly both with $u$ and $R$. The cavity
deformation threshold is defined by the condition $p_{\rm min} = 0$ and is reached easier for larger projectiles, at the
critical velocity $u_{\rm cr}\simeq2p_{\rm st}/(\rho\nu R)$ [here we assume $p_{\rm st}\lesssim \rho(\nu R)^2$/8]. Note
that in the high-velocity regime $u > 2\nu R/9$ the pressure minimum is shifted to a certain latitude in the rear
hemisphere, approaching the ``equator'' ($\xi=0$) asymptotically. In this case friction plays a minor role and the
critical velocity is $u_{\rm cr}\simeq\sqrt{8 p_{\mbox{\scriptsize st}}/5 \rho}$.

Note that for arbitrary acceleration (assuming $\dot{\bf u}\|{\bf u}$), one can introduce the effective damping rate
$\nu_{\rm eff}=\nu+\dot u/u$ which can have either sign, so that the critical cosine $\xi_{\rm cr}$ can have either sign as
well. Thus, if a projectile experiences sufficiently strong deceleration ($|\dot u|/u>\nu$), the cavity can also be deformed
at the front hemisphere.

The critical velocity $u_{\mbox{\scriptsize cr}}$ 
calculated above defines the emergence of a point or circle of 
zero pressure on the cavity surface depending on the value 
of $\xi _{\mbox{\scriptsize cr}} $. It is well-known in gas 
dynamics that the existence of small local regions of negative 
pressure would not lead to a significant deformation of flow 
lines and, therefore, to a stall. Instead, stall emergence 
requires an extended region of negative pressure that covers 
a significant part of ``rear hemisphere''. There does not seem to be any 
criterion that would tell us how large 
the negative region should be to initiate a stall. 
Nevertheless, one can hope that the velocity 
$u_{\mbox{\scriptsize cr}}$ derived above provides a 
reasonable lower bound estimate for the cavity deformation 
onset. Strictly speaking, the theory discussed above is not 
applicable for the case $u > u_{\mbox{\scriptsize cr}}$ 
because a streamlined cavity is 
already deformed. At the same time below the stall 
threshold, this deformation 
is small, i.e., the cavity is still almost spherical. 
Therefore, one could extend the analysis to the region 
of the velocities higher than the critical one and, 
respectively, to the region of negative 
pressure larger than a point or a circle on 
the cavity surface. In the next Section, 
obtained results are generalized to the case of the finite 
compressibility of a fluid.

\section{\label{s3} FLOW WITH FINITE COMPRESSIBILITY}

In this Section, we will estimate the effect of finite compressibility of the dust cloud on the velocity and pressure
fields. Here, we will treat a steady ($\dot{\bf u} = 0$) and irrotational ($\nabla \times {\bf{v}} = 0$) flow, so that the
fluid density is a function solely of the radius-vector: $\rho = \rho ({\bf{r}})$. Since $\nabla \times {\bf{v}} = 0$ and
$\nabla p = c^2 \nabla \rho $, where $c = \sqrt {(\partial p/\partial \rho )_T }$ is the sound velocity, we can rewrite
Eq.~(\ref{e1}) in the form
\begin{equation}
({\bf{v}}\cdot\nabla ){\bf{v}} + \nu ({\bf{v}} + {\bf{u}})
= - \frac{{c^2 }}{\rho }\nabla \rho . \label{e14}
\end{equation}
In this case, the continuity equation includes the fluid
density,
\begin{equation}
\nabla \cdot {\bf{v}} + {\bf{v}} \cdot \frac{{\nabla \rho
}}{\rho } = 0. \label{e15}
\end{equation}
Here, $c$ is assumed to be constant independent of the local velocity and plasma state parameters. We substitute $\nabla
\rho /\rho = - (1/c^2 )({\bf{v}}\cdot\nabla ){\bf{v}} - (\nu /c^2 )({\bf{v}} + {\bf{u}})$
 from (\ref{e14}) in (\ref{e15}) to derive
\begin{equation}
\nabla \cdot {\bf{v}} = \frac{1}{{c^2 }}{\bf{v}} \cdot
({\bf{v}} \cdot \nabla ){\bf{v}} + \frac{\nu }{{c^2 }}(v^2
+ {\bf{u}} \cdot {\bf{v}}). \label{e16}
\end{equation}

We will search for the solution of Eq.~(\ref{e16}) in the form ${\bf{v}} = {\bf{v}}_0 + {\bf{v}}_1 $, where ${\bf{v}}_0({\bf
r })$ is the solution of incompressible problem (for $\rho = \rho _0 =$~const, i.e., for $c\to\infty$) given by
Eq.~(\ref{e5}), the respective pressure distribution $p_0({\bf r })$ is presented by Eq.~(\ref{e8}) (for $\dot u=0$).
Obviously, $\nabla \cdot {\bf{v}}_0 = 0$. We assume that $\left| {{\bf{v}}_1 } \right| \ll \left| {{\bf{v}}_0 } \right|$ and
retain solely zeroth-order terms on the rhs of (\ref{e16}) to obtain
\begin{equation}
\nabla \cdot {\bf{v}}_1 = \frac{1}{{2c^2 }}{\bf{v}}_0
\cdot \nabla v_0^2 + \frac{\nu }{{c^2 }}(v_0^2 + {\bf{u}}
\cdot {\bf{v}}_0 ). \label{e18}
\end{equation}
Here, we took into account that $\nabla \times {\bf v}_0 = 0$ and hence $({\bf{v}}_0 \cdot \nabla ){\bf{v}}_0 = \nabla v_0^2
/2$. Since $\left| {{\bf{v}}_0 } \right| \sim u$, $\left| \nabla \right| \sim 1/R$, and $(u/c)^2$ is small for subsonic
motion, the estimate $\left| {{\bf{v}}_1 } \right| \sim (u^2 /c^2 )\max \{ u,\,\nu R\}$ following from (\ref{e18}) justifies
our assumption. It is seen from this estimate that both sides of Eq.~(\ref{e18}) are of the order $(u/c)^2 $, and terms
other that zeroth-order ones would lead to higher-order terms in this parameter. Obviously, $\left| {{\bf{v}}_1 } \right|
\to 0$ at $c \to \infty $.

We introduce the scaled coordinate $x = r/R$ and the dimensionless potential $\tilde \varphi_1$ defined by the relation
${\bf{v}}_1 = (u^3 /c^2 )\nabla \tilde \varphi_1$ (in so doing, the condition $\nabla \times {\bf{v}}_1=0$ is satisfied) to
rewrite (\ref{e18}) in the dimensionless form
\begin{eqnarray}\label{e19}
{\nabla^2\tilde \varphi_1=\frac{\kappa}{{2x^6}}P_0(\xi)+\left(\frac{36}{5x^7}-\frac{9}{{2x^{10}}}\right)P_1(\xi)}\nonumber \\
{-\kappa\left(\frac1{x^3}-\frac1{2x^6}\right)P_2(\xi)}\nonumber \\ {-\left(\frac6{x^4}-\frac{24}{5x^7}+\frac3{2x^{10}}\right)P_3(\xi),} 
\end{eqnarray}
where $P_l (\xi )$ are the Legendre polynomials and $\kappa = \nu R/u$. Here and in what follows, differentiation with
respect to scaled coordinates is assumed. The boundary conditions for Eq.~(\ref{e19}) follow from the conditions ${\bf{v}}_1
(\infty )= 0$ and ${\bf{n}} \cdot {\bf{v}}_1 \left| {_{r = R} } \right. =0$, i.e.,
\begin{equation}
\tilde \varphi_1 (\infty ) = 0,\quad \left.\frac{d\tilde \varphi_1}{d{\bf n }}\right|_{x = 1} = 0.
\label{e20}
\end{equation}

Solution of Eq.~(\ref{e19}) is derived in Appendix \ref{sA}. The resulting finite-compressibility corrections to the
potential and velocity at the cavity surface follow from (\ref{e29}):
\begin{equation}
\begin{array}{*{20}c}
  {\tilde\varphi_1\left|_{x = 1}\right.=\frac{81}{176}\xi^3+\frac{\kappa}2\left(\frac14\xi^2-\frac13\right)-\frac{551}{880},} \\
  {{\bf{v}}_1\left|_{x=1}\right.=\frac{u^2 }{c^2}\left(\frac{243}{176}\xi^2+\frac{\kappa}4\xi-\frac{551}{880}\right)({\bf{u}}
  -\xi u{\bf{n}}).} \\
\end{array} \label{e21}
\end{equation}
As it could be expected from symmetry considerations, ${\bf{v}}_1 \left| {_{x = 1} } \right. = 0$ at $\xi =\pm1$, i.e, at
both poles. Note that result (\ref{e21}) validates the estimate $\left| {{\bf{v}}_1 } \right| \sim (u^2 /c^2 )\max \{
u,\,\nu R\}$ given above.

On the basis of velocity correction (\ref{e21}) and using Eq.~(\ref{e14}), it is a straightforward matter to calculate a
compressibility correction to the pressure $p_1 = p - p_0$. This is performed in Appendix \ref{sB}. It is worth mentioning
that in the general case, correction to the pressure (\ref{e33}) includes solely even powers of $\xi $, so that
corresponding correction to an additional force exerted on the projectile due to dissipation in the surrounding particle
fluid [Eq.~(10) in Ref.~\cite{25}] vanishes. In the friction-dominated regime (large $\kappa$), one can retain in
(\ref{e33}) solely the term proportional to $u^2 $, which is independent of $\xi $, so that the total pressure is
\begin{equation}
p(\xi ) \simeq p_{\rm st}+\frac{\rho_0u\nu R}2\xi+\frac{\rho_0u^2}8\left(9\xi^2-5+\frac{4\nu^2R^2}{3c^2} \right).
\label{e22}
\end{equation}
Thus, the pressure increases due to finite compressibility, 
so that the threshold velocity of cavity deformation 
increases as well. On the other hand, one can see that the 
resulting correction is small when $\kappa u/c \ll 1$. 
However, we will see that the finite compressibility 
correction is always small for large projectiles even if this 
condition is not satisfied.

Now we repeat the analysis of Sec.~\ref{s2} for 
Eq.~(\ref{e22}) instead of (\ref{e8}). Since the finite 
compressibility correction is independent of $\xi $, the 
expression for $\xi _{\mbox{\scriptsize cr}}$ remains the 
same as in Sec.~\ref{s2}. Consider first the case $\left| {\xi 
_{\mbox{\scriptsize cr}} } \right| > 1$, when we have to set 
$\xi = - 1$
 in Eq.~(\ref{e22}) to find the threshold velocity 
corresponding to emergence of a point of zero pressure at 
the ``rear pole" of a cavity:
\begin{equation}
u_{\mbox{\scriptsize cr}} = \frac{{\nu R}}{{2\omega }} - \sqrt {\frac{{\nu ^2 R^2 }}{{4\omega ^2 }} - \frac{{2p_{\mbox{\scriptsize st}} }}{{\omega \rho _0 }}} \simeq \frac{{2p_{\mbox{\scriptsize st}} }}{{\nu R\rho _0 }}, \label{e001}
\end{equation}
where $\omega = 1 + \nu ^2 R^2 /3c^2 $. For the case of 
large cavities ($R \to \infty $) typical for treated case, 
approximation (\ref{e001}) is valid provided that the 
condition $p_{\mbox{\scriptsize st}} /\rho _0 c^2 \ll 1$, 
which applies to experiments analyzed in Sec.~\ref{s4}, is 
satisfied even if $\kappa u/c > 1$. Thus, finite 
compressibility seems to have no significant effect on the 
critical velocity. The reciprocal dependence
$p_{\mbox{\scriptsize st}}( u_{\mbox{\scriptsize cr}})$ has the form
\begin{equation}
\frac{{p_{\mbox{\scriptsize st}} 
}}{{\rho _{\mbox{\scriptsize 0}}}}
 = \frac{{\omega u_{\mbox{\scriptsize cr}}^2 }}{2}\left( 
{\frac{{\nu R}}{{\omega u_{\mbox{\scriptsize cr}} }} - 
1} \right) \simeq \frac{1}{2}\nu 
Ru_{\mbox{\scriptsize cr}} .
\label{e040}
\end{equation}

Upon further increase of the projectile velocity $u$, the region of negative pressure on the surface of a cavity is expanded. It is a straightforward matter to demonstrate that extrapolation of the theory to the extended region of negative pressures (where it is formally invalid) results in a very fast increase of this region with the increase of $u$. Therefore, $u_{\mbox{\scriptsize cr}}$ is likely to be a good estimate for the stall velocity $u_{\mbox{\scriptsize st}}$. At the same time, $u_{\mbox{\scriptsize st}} > u_{\mbox{\scriptsize cr}}$ means that if we equate $u_{\mbox{\scriptsize cr}} = u_{\mbox{\scriptsize st}}$ we obtain from  (\ref{e040}) an overestimated $p_{\mbox{\scriptsize st}}$, which can be treated as {\it an upper bound estimate\/} for the static pressure. 

In the case $\left| {\xi _{\mbox{\scriptsize cr}} } \right| < 
1$
 typical for small projectiles, a circle with zero pressure first 
emerges at the critical velocity
\begin{eqnarray}\label{e003}
u_{\mbox{\scriptsize cr}} = 2\left( 
{\frac{{2p_{\mbox{\scriptsize st}} }}{{5\rho _0 }} - 
\frac{{\nu ^2 R^2 }}{{45}}} \right)^{1/2} \left( {1 - 
\frac{{4\nu ^2 R^2 }}{{15c^2 }}} \right)^{ - 1/2}\nonumber \\
\simeq 
\left( {\frac{{8p_{\mbox{\scriptsize st}} }}{{5\rho _0 }}} 
\right)^{1/2} . \quad
\end{eqnarray}
Here, the approximation corresponding to the limit
$R \to 0$
also coincides with that for an incompressible fluid. If 
$u_{\mbox{\scriptsize cr}}$ is known from experiment, 
one can derive an upper bound estimate for the static 
pressure,
\begin{equation}
\frac{{p_{\mbox{\scriptsize st}} }}{{\rho _0 }} = \frac{{5u_{\mbox{\scriptsize cr}}^2 }}{8} + \frac{{\nu ^2 R^2 }}{{18}}\left( {1 - \frac{{3u_{\mbox{\scriptsize cr}}^2 }}{{c^2 }}} \right) \simeq \frac{{5u_{\mbox{\scriptsize cr}}^2 }}{8}. \label{e004}
\end{equation}
In this case at $u > u_{\mbox{\scriptsize cr}}$, the circle 
expands to a region between two circles on the surface of a cavity.
For small projectiles, one can also demonstrate that a fast 
expansion of the negative region takes place as $u$
is increased. Thus, $u_{\mbox{\scriptsize cr}}$ also yields a 
reasonable order of magnitude estimate for a stall threshold.

\section{\label{s4} COMPARISON WITH 
EXPERIMENT}

Based on developed theory we will analyze the possibility 
of cavity deformation in experiments. Consider first the 
experiments carried out in the PK-3 Plus Laboratory 
onboard the ISS under microgravity conditions. Details on 
the setup can be found in \cite{18}. Dust particles injected 
into the main plasma with dispensers formed a cloud 
around the center of the chamber. Some larger particles 
present in the chamber as well get sporadically accelerated 
and penetrate into the cloud, thus forming projectiles 
\cite{26}. Consider the experiment with argon as a carrier 
gas. For a gas pressure and temperature of 10 Pa and 
$T_n = 300\;{\mbox{K}}$, respectively, we have $m_n = 
6.63 \times 10^{ - 23} \;{\mbox{g}}$, $n_n = 2.42 \times 
10^{15} \,{\mbox{cm}}^{ - 3} $, and $v_{T_n } = 2.5 
\times 10^4 \;{\mbox{cm/s}}$. A dust cloud was formed by 
melamine-formaldehyde particles with the radius $a_d = 
1.275 \times 10^{ - 4} \;{\mbox{cm}}$
 and mass $M_d = 1.31 \times 10^{ - 11} \;{\mbox{g}}$
 \cite{19,22}, therefore, $\nu = 46.6\;{\mbox{s}}^{ - 1} $.

\begin{figure}
\unitlength=0.24pt
\begin{picture}(500,750)
\put(-490,-150){\includegraphics[width=4.5in]{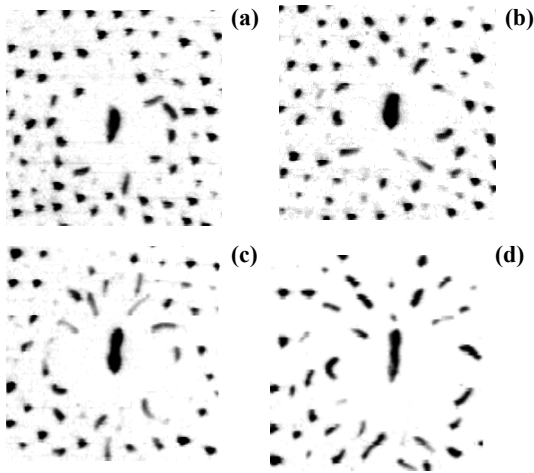}}
\end{picture}
\caption{\label{f4}Positions of dust particles and the projectile 
(negative images) at successive instants. Time interval 
between individual frames is 0.08 s. Frames (a) and (b) 
illustrate a spherical cavity; (c) and (d), deformed cavity. 
The projectile is accelerated from (a) to (d).}
\end{figure}

Successive frames of the path of a projectile recorded by a 
high-resolution camera are shown in Fig.~\ref{f4}. 
The 
projectile velocity increases slowly from $u = 
0.7\;{\mbox{cm/s}}$
 to $1.4\;{\mbox{cm/s}}$. Since the sound velocity 
measured in this experiment amounts to 2.2 cm/s \cite{19}, 
such motion is subsonic. The peculiarity of this motion is 
an abrupt change of cavity configuration. At the beginning, 
the projectile moves in the center of a cavity 
[cf.~Figs.~\ref{f4}(a) and (b)]. Then, at the point of trajectory 
characterized by the velocity $u = 1.06\;{\mbox{cm/s}}$, 
acceleration $\dot u = 2.6\;{\mbox{cm/s}}^{2} $, 
and the cavity radius $R = 3.74 \times 10^{ - 2} 
\;{\mbox{cm}}$,
 the projectile shifts abruptly from the cavity center to its front 
side [Fig.~\ref{f4}(c)]. In the course of further motion, 
cavity deformation is preserved [Fig.~\ref{f4}(d)]. One can 
assume that at the point shown in Fig.~\ref{f4}(c), the 
velocity of the accelerating projectile exceeds a stall threshold. 
Under specified experimental conditions, $\xi _{\mbox{\scriptsize cr}} = -0.365$, 
hence we substitute $u_{\mbox{\scriptsize cr}} = 
1.06\;{\mbox{cm/s}}$
in (\ref{e004}) to derive an upper bound estimate 
$p_{\mbox{\scriptsize st}} /\rho _0 \approx 
0.753\;{\mbox{cm}}^2 {\mbox{/s}}^2$ (or 
$p_{\mbox{\scriptsize st}} \approx 3.0 \times 10^{ - 7} 
\;{\mbox{Pa}}$). 
Note that the correction for acceleration to 
the total pressure $ - \rho \dot uR/2$
 [Eq.~(\ref{e8})] is small due to a small projectile 
acceleration at selected point ($\dot u/u \ll \nu$) and can be neglected. 

The estimate obtained above can be compared with a 
theoretical estimation for the static pressure, which is based 
on dimensionality considerations. Since dimensionalities of 
the pressure and energy density coincide, we can write with 
an order of magnitude accuracy $p_{\mbox{\scriptsize st}} 
\sim Z_d^2 e^2 n_d /2\bar r$, where $\bar r = (3/4\pi n_d 
)^{1/3}$ is the Wigner--Seitz radius for a dust crystal
and $n_d$ is the number density of dust particles. With 
$Z_d = 1200$
and $n_d = 3 \times 10^5 \;{\mbox{cm}}^{ - 3}$ 
\cite{19}, this estimation yields
$p_{\mbox{\scriptsize st}} \sim 5.4 \times 10^{ - 7} \;{\mbox{Pa}}$, 
which correlates 
with the upper bound estimate resulting from experiment.

\begin{figure}
\unitlength=0.24pt
\begin{picture}(500,750)
\put(-475,-240){\includegraphics[width=4.9in]{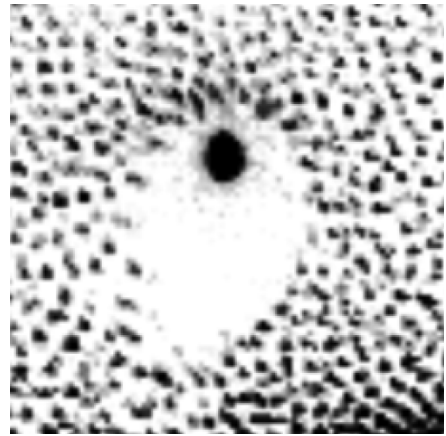}}
\end{picture}
\caption{\label{f5}Deformed cavity around a large subsonic 
projectile moving with the velocity $u = 
1.8\;{\mbox{cm/s}}$. Reproduction of an enlarged 
fragment of Fig.~3a in Ref.~\cite{11}.}
\end{figure}

Microgravity conditions can also be maintained on 
parabolic flights. Such experiments were performed using 
the IMPF-K2 chamber \cite{11}. The experiment was 
carried out with argon at 30 Pa and $a_d = 4.775 \times 
10^{ - 4} \;{\mbox{cm}}$
 ($M_d = 6.88 \times 10^{ - 10} \;{\mbox{g}}$
 and $\nu = 33.5\;{\mbox{s}}^{ - 1} $). Projectiles were 
accelerated by a special device (cogwheel) up to both 
supersonic and subsonic velocities. A typical image 
illustrating the motion of a subsonic projectile inside a dust 
cloud is shown in Fig.~\ref{f5}.
Both the projectile shift 
from the cavity center toward its front side and the nonspherical 
(elongated) shape of a cavity are visible. Unfortunately, a 
point of transition from a spherical to a deformed cavity was 
not detected in this experiment. Nevertheless, we can 
estimate the stall threshold for this case. Note that here, the 
estimate $p_{\mbox{\scriptsize st}} \sim Z_d^2 e^2 n_d 
/2\bar r$
 with $Z_d = 8800$
 and $n_d = 2 \times 10^4 \;{\mbox{cm}}^{ - 3}$ 
\cite{11} yields $p_{\mbox{\scriptsize st}} \sim 7.8 \times 
10^{ - 7} \;{\mbox{Pa}}$, which is close to the estimate 
for experiment \cite{22} discussed above. Therefore, we 
can assume that $p_{\mbox{\scriptsize st}}$ are 
close in both systems and use the upper bound 
estimate (\ref{e001}) to calculate 
$u_{\mbox{\scriptsize cr}}$ for the treated 
system. With the cavity radius roughly estimated as $R = 
0.1\;{\mbox{cm}}$, we derive $u_{\mbox{\scriptsize cr}} \approx 
0.462\;{\mbox{cm/s}}$, so that the projectile moves 
in the low-velocity regime ($\xi _{\mbox{\scriptsize cr}} = -1.61$).
Therefore, the stall threshold for 
this system does not exceed 1~cm/s. This result correlates 
with the projectile velocity in Fig.~\ref{f5}, $u = 
1.8\;{\mbox{cm/s}}$, which is still lower than the sound 
velocity $c = 2.0\;{\mbox{cm/s}}$
 \cite{11}. Thus, the main notion of the present study, namely that the 
cavity around a moving projectile can be deformed at {\it 
subsonic velocity\/}, is corroborated.

\section{\label{s5} CONCLUSION}

We have investigated the possibility of the deformation of a cavity around large projectiles slowly moving inside a cloud of
small dust particles. This cavity is formed due to a strong Coulomb repulsion between the projectile and dust particles.
Since the parameter $\beta_{dp}$ characterizing this interaction is very large, cavity deformation occurs solely due to
the formation of a void adjacent to the initially spherical cavity. This is similar to the process of void formation in a fluid
under negative pressure. Since we approximate the dust cloud by a system of hard spheres, zero pressure is sufficient for
void formation leading to cavity deformation.

We model a collective particle subsonic motion in a complex plasma by nonviscous irrotational hydrodynamic motion of dust
fluid and show that, in this model, the reason for cavity deformation is the friction between dust particles and neutral
atoms of a carrier gas. It is this friction that stipulates the emergence of a zone of negative pressure over the back of a
cavity surface. If such a zone covers a substantial part of the surface, stall takes place, which is observed as cavity
deformation. We have found that the deformation is more likely to be observed for larger projectiles. The deformation occurs
when the projectile velocity exceeds some threshold of subsonic velocity.

Since typical projectile velocities are of the same order of magnitude as the sound velocity, we have studied the effect of
finite compressibility of the dust fluid on velocity and pressure fields. Toward that end, we have obtained an exact
solution of linearized gas dynamics equations for compressible nonviscous fluid flowing about the cavity. In so doing, we
included both the convective and friction terms in the Navier--Stokes equation. Calculations showed that corrections for
finite compressibility are rather small. Generally, account of the compressibility leads to a shift of the area
covered by a negative pressure zone.

Analysis of available experimental data has validated our approach. It was found that the cavity is deformed when either the
velocity \cite{22} or the size \cite{11} of a projectile is sufficiently large. It is worth mentioning that the phenomenon
of cavity deformation is similar to the formation of a void behind a cluster of smaller particles moving in the cloud of
larger ones \cite{27}. For intermediate projectile sizes, a transition from a spherical to a deformed cavity was observed along
the trajectory of an individual projectile. The transition threshold is in accordance with the developed theory. This made
it possible to estimate the static pressure inside a dust cloud and, hence, opened up a possibility to use projectiles as a
diagnostic tool for the dust equation of state.

In this study, we restricted ourselves to an investigation of the threshold of cavity deformation and did not address the shape
of the deformed cavity or displacement of the projectile away from the center of its cavity. This problem 
would require an adequate account of the local nonspherical charge distribution around the projectile. It is also possible that the observed shift of the projectile from the center of a cavity is noticeably favored by an additional drag force exerted on the projectile \cite{25}.

\appendix
\section{\label{sA}}

The solution of Eq.~(\ref{e19}) with boundary conditions (\ref{e20}) is represented in the form
\begin{equation}
\tilde \varphi_1 (x,\,\xi ) = \sum\limits_{l = 0}^3 f_l (x)P_l(\xi ) , \label{e24}
\end{equation}
where $f_l (x)$ are the coefficients of expansion in the Legendre polynomials $P_l (\xi )$. We substitute (\ref{e24}) in
(\ref{e19}) taking into account that $\nabla ^2 P_l (\xi ) = - l(l + 1)x^{-2}P_l (\xi )$ to reduce Eq.~(\ref{e19}) to a set
of ordinary differential equations in spherical coordinates for nonzero functions $f_l (x)$,
\begin{equation}
\frac{d}{{dx}}\left( {x^2 \frac{{df_l }}{{dx}}} \right) -l(l+ 1)f_l = q_l (x), \label{e25}
\end{equation}
where $q_0 (x) = \frac12\kappa x^{-4} $, $q_1 (x) =\frac{36}5x^{-5}-\frac92x^{-8}$, $q_2 (x) = -\kappa x^{-1}+\frac12\kappa
x^{-4}$, $q_3 (x) =-6x^{-2}+\frac{24}5x^{-5}-\frac32x^{-8}$. The boundary conditions for (\ref{e25}) follow from
(\ref{e20}):
\begin{equation}
f_l (\infty ) = 0,\quad \left.\frac{df_l }{dx}\right|_{x =1} = 0. \label{e26}
\end{equation}
The rhs of (\ref{e25}) has the form $\sum\limits_k b_{lk}x^{-k}$, where $b_{lk}$ are constants entering the functions
$q_l(x)$. Solution of (\ref{e25}) is then given by
\begin{equation}
f_l (x) = \frac{C}{{x^{l + 1} }} + \sum\limits_k {\frac{{b_{lk} }}{{k(k - 1) - l(l + 1)}}\frac{1}{{x^k }}} , \label{e27}
\end{equation}
where $C$ is defined by the second boundary condition (\ref{e26}), whence it follows that
\begin{equation}
f_l (x) = \sum\limits_k {\frac{{b_{lk} }}{{l(l + 1) - k(k -
1)}}\left( {\frac{k}{{l + 1}}\frac{1}{{x^{l + 1} }} -
\frac{1}{{x^k }}} \right)} . \label{e28}
\end{equation}
Finally, we obtain
\begin{eqnarray}\label{e29}
  {f_0 = \kappa\left(-\frac1{6x}+\frac1{24x^4}\right),}\\
  {f_1 = -\frac2{3x^2}+ \frac2{5x^5}-\frac1{12x^8 },} \\
  {f_2 = \kappa\left(\frac1{6x}-\frac1{6x^3}+\frac1{12x^4}\right),} \\
  {f_3 = \frac3{5x^2}-\frac{54}{55x^4}+\frac3{5x^5}-\frac3{88x^8}.} \\
\end{eqnarray} 

\section{\label{sB}}

The pressure correction due to finite compressibility can be found using the corresponding velocity correction,
Eq.~(\ref{e21}). We rewrite Eq.~(\ref{e14}) in the form
\begin{equation}
 - c^2 \nabla \ln \rho = \frac{1}{2}\nabla v^2 + \nu \nabla
\varphi , \label{e30}
\end{equation}
where $\varphi = \varphi _0 + \varphi _1$ and ${\bf{v}}={\bf u}+\nabla \varphi $. By integrating Eq.~(\ref{e30}) we derive
\begin{eqnarray}\label{e31}
\rho - \rho _0 = - \frac{{\rho _0 }}{{c^2 }}\left(
{\frac{{v_0^2 }}{2} + \nu \varphi _0 } \right) - \frac{{\rho
_0 }}{{c^2 }}\left( {{\bf{v}}_0 \cdot {\bf{v}}_1 + \nu
\varphi _1 } \right) \nonumber \\ + \frac{{\rho _0 }}{{2c^4 }}\left(
{\frac{{v_0^2 }}{2} + \nu \varphi _0 } \right)^2 ,\quad
\end{eqnarray}
where $\rho _0 = \rho \left| {_{r = \infty } } \right.$ and the terms up to $(u/c)^4 $ are retained (for simplicity we
assume $c =$~const). We write the equation of state for the dust fluid in the form $p - p_{\rm st} = c^2 (\rho - \rho _0 )$,
which yields
\begin{eqnarray}\label{e32}
{p = p_{\rm st} - \frac{{\rho _0 v_0^2 }}{2} - \nu \rho _0
\varphi _0 - \rho _0 {\bf{v}}_0 \cdot {\bf{v}}_1 - \nu \rho
_0 \varphi _1  \nonumber } \\ { + \frac{1}{{2\rho _0 c^2 }}\left( {\frac{{\rho
_0 v_0^2 }}{2} + \nu \rho _0 \varphi _0 } \right)^2 .} \\ \nonumber \quad
\end{eqnarray}
The first three terms of Eq.~(\ref{e32}) represent Eq.~(\ref{e8}) with $\dot u = 0$. Hence, by substituting in
Eq.~(\ref{e32}) the expressions for $\varphi_0$ and ${\bf v}_0$ [Eqs.~(\ref{e4}) and (\ref{e5})] as well as for $\varphi_1$
and ${\bf v}_1$ [Eq.~(\ref{e21})] taken  at the cavity surface, we get
\begin{eqnarray}
p_1 (\xi )=\frac{\rho _0 u^4}{c^2}\left[\frac16\kappa^2-3\kappa\xi^2\left(\xi^2-\frac{101}{440}\right)\right.\nonumber\\
\left.-\frac{2025}{1408}\left(\xi^4-\frac{5414}{3375}\xi^2+\frac{5237}{10125}\right)\right]. \label{e33}
\end{eqnarray}

\providecommand{\noopsort}[1]{}\providecommand{\singleletter}[1]{#1}%


\begin{thebibliography}{27}%
\makeatletter
\providecommand \@ifxundefined [1]{%
 \@ifx{#1\undefined}
}%
\providecommand \@ifnum [1]{%
 \ifnum #1\expandafter \@firstoftwo
 \else \expandafter \@secondoftwo
 \fi
}%
\providecommand \@ifx [1]{%
 \ifx #1\expandafter \@firstoftwo
 \else \expandafter \@secondoftwo
 \fi
}%
\providecommand \natexlab [1]{#1}%
\providecommand \enquote  [1]{``#1''}%
\providecommand \bibnamefont  [1]{#1}%
\providecommand \bibfnamefont [1]{#1}%
\providecommand \citenamefont [1]{#1}%
\providecommand \href@noop [0]{\@secondoftwo}%
\providecommand \href [0]{\begingroup \@sanitize@url \@href}%
\providecommand \@href[1]{\@@startlink{#1}\@@href}%
\providecommand \@@href[1]{\endgroup#1\@@endlink}%
\providecommand \@sanitize@url [0]{\catcode `\\12\catcode `\$12\catcode
  `\&12\catcode `\#12\catcode `\^12\catcode `\_12\catcode `\%12\relax}%
\providecommand \@@startlink[1]{}%
\providecommand \@@endlink[0]{}%
\providecommand \url  [0]{\begingroup\@sanitize@url \@url }%
\providecommand \@url [1]{\endgroup\@href {#1}{\urlprefix }}%
\providecommand \urlprefix  [0]{URL }%
\providecommand \Eprint [0]{\href }%
\providecommand \doibase [0]{http://dx.doi.org/}%
\providecommand \selectlanguage [0]{\@gobble}%
\providecommand \bibinfo  [0]{\@secondoftwo}%
\providecommand \bibfield  [0]{\@secondoftwo}%
\providecommand \translation [1]{[#1]}%
\providecommand \BibitemOpen [0]{}%
\providecommand \bibitemStop [0]{}%
\providecommand \bibitemNoStop [0]{.\EOS\space}%
\providecommand \EOS [0]{\spacefactor3000\relax}%
\providecommand \BibitemShut  [1]{\csname bibitem#1\endcsname}%
\let\auto@bib@innerbib\@empty
\bibitem [{\citenamefont {Fortov}\ and\ \citenamefont {Morfill}(2009)}]{1}%
  \BibitemOpen
  \bibinfo {editor} {\bibfnamefont {V.~E.}\ \bibnamefont {Fortov}}\ and\
  \bibinfo {editor} {\bibfnamefont {G.~E.}\ \bibnamefont {Morfill}},\ eds.,\
  \href@noop {} {\emph {\bibinfo {title} {Complex and Dusty Plasmas: From
  Laboratory to Space}}},\ Series in Plasma Physics\ (\bibinfo  {publisher}
  {CRC Press},\ \bibinfo {year} {2009})\BibitemShut {NoStop}%
\bibitem [{\citenamefont {Chu}\ and\ \citenamefont {I}(1994)}]{2}%
  \BibitemOpen
  \bibfield  {author} {\bibinfo {author} {\bibfnamefont {J.~H.}\ \bibnamefont
  {Chu}}\ and\ \bibinfo {author} {\bibfnamefont {L.}~\bibnamefont {I}},\
  }\href@noop {} {\bibfield  {journal} {\bibinfo  {journal} {Phys.\ Rev.\
  Lett.}\ }\textbf {\bibinfo {volume} {72}},\ \bibinfo {pages} {4009} (\bibinfo
  {year} {1994})}\BibitemShut {NoStop}%
\bibitem [{\citenamefont {Thomas}\ \emph {et~al.}(1994)\citenamefont {Thomas},
  \citenamefont {Morfill}, \citenamefont {Demmel}, \citenamefont {Goree},
  \citenamefont {Feuerbacher},\ and\ \citenamefont {M{\"{o}}hlmann}}]{3}%
  \BibitemOpen
  \bibfield  {author} {\bibinfo {author} {\bibfnamefont {H.}~\bibnamefont
  {Thomas}}, \bibinfo {author} {\bibfnamefont {G.~E.}\ \bibnamefont {Morfill}},
  \bibinfo {author} {\bibfnamefont {V.}~\bibnamefont {Demmel}}, \bibinfo
  {author} {\bibfnamefont {J.}~\bibnamefont {Goree}}, \bibinfo {author}
  {\bibfnamefont {B.}~\bibnamefont {Feuerbacher}}, \ and\ \bibinfo {author}
  {\bibfnamefont {D.}~\bibnamefont {M{\"{o}}hlmann}},\ }\href@noop {}
  {\bibfield  {journal} {\bibinfo  {journal} {Phys.\ Rev.\ Lett.}\ }\textbf
  {\bibinfo {volume} {73}},\ \bibinfo {pages} {652} (\bibinfo {year}
  {1994})}\BibitemShut {NoStop}%
\bibitem [{\citenamefont {Hayashi}\ and\ \citenamefont {Tashibana}(1994)}]{4}%
  \BibitemOpen
  \bibfield  {author} {\bibinfo {author} {\bibfnamefont {Y.}~\bibnamefont
  {Hayashi}}\ and\ \bibinfo {author} {\bibfnamefont {S.}~\bibnamefont
  {Tashibana}},\ }\href@noop {} {\bibfield  {journal} {\bibinfo  {journal}
  {Jpn.\ J.\ Appl.\ Phys.}\ }\textbf {\bibinfo {volume} {33}},\ \bibinfo
  {pages} {L804} (\bibinfo {year} {1994})}\BibitemShut {NoStop}%
\bibitem [{\citenamefont {Vladimirov}\ \emph {et~al.}(2005)\citenamefont
  {Vladimirov}, \citenamefont {Ostrikov},\ and\ \citenamefont {Samarian}}]{5}%
  \BibitemOpen
  \bibfield  {author} {\bibinfo {author} {\bibfnamefont {S.~V.}\ \bibnamefont
  {Vladimirov}}, \bibinfo {author} {\bibfnamefont {K.}~\bibnamefont
  {Ostrikov}}, \ and\ \bibinfo {author} {\bibfnamefont {A.~A.}\ \bibnamefont
  {Samarian}},\ }\href@noop {} {\emph {\bibinfo {title} {Physics and
  applications of complex plasmas}}}\ (\bibinfo  {publisher} {Imperial
  College},\ \bibinfo {address} {London},\ \bibinfo {year} {2005})\BibitemShut
  {NoStop}%
\bibitem [{\citenamefont {Fortov}\ \emph
  {et~al.}(2005{\natexlab{a}})\citenamefont {Fortov}, \citenamefont {Ivlev},
  \citenamefont {Khrapak}, \citenamefont {Khrapak},\ and\ \citenamefont
  {Morfill}}]{6}%
  \BibitemOpen
  \bibfield  {author} {\bibinfo {author} {\bibfnamefont {V.}~\bibnamefont
  {Fortov}}, \bibinfo {author} {\bibfnamefont {A.}~\bibnamefont {Ivlev}},
  \bibinfo {author} {\bibfnamefont {S.}~\bibnamefont {Khrapak}}, \bibinfo
  {author} {\bibfnamefont {A.}~\bibnamefont {Khrapak}}, \ and\ \bibinfo
  {author} {\bibfnamefont {G.}~\bibnamefont {Morfill}},\ }\href@noop {}
  {\bibfield  {journal} {\bibinfo  {journal} {Phys.\ Rep.}\ }\textbf {\bibinfo
  {volume} {421}},\ \bibinfo {pages} {1} (\bibinfo {year}
  {2005}{\natexlab{a}})}\BibitemShut {NoStop}%
\bibitem [{\citenamefont {Morfill}\ and\ \citenamefont {Ivlev}(2009)}]{7}%
  \BibitemOpen
  \bibfield  {author} {\bibinfo {author} {\bibfnamefont {G.~E.}\ \bibnamefont
  {Morfill}}\ and\ \bibinfo {author} {\bibfnamefont {A.~V.}\ \bibnamefont
  {Ivlev}},\ }\href@noop {} {\bibfield  {journal} {\bibinfo  {journal} {Rev.\
  Mod.\ Phys.}\ }\textbf {\bibinfo {volume} {81}},\ \bibinfo {pages} {1353}
  (\bibinfo {year} {2009})}\BibitemShut {NoStop}%
\bibitem [{\citenamefont {Shukla}\ and\ \citenamefont {Eliasson}(2009)}]{8}%
  \BibitemOpen
  \bibfield  {author} {\bibinfo {author} {\bibfnamefont {P.~K.}\ \bibnamefont
  {Shukla}}\ and\ \bibinfo {author} {\bibfnamefont {B.}~\bibnamefont
  {Eliasson}},\ }\href@noop {} {\bibfield  {journal} {\bibinfo  {journal}
  {Rev.\ Mod.\ Phys.}\ }\textbf {\bibinfo {volume} {81}},\ \bibinfo {pages}
  {25} (\bibinfo {year} {2009})}\BibitemShut {NoStop}%
\bibitem [{\citenamefont {Bonitz}\ \emph {et~al.}(2010)\citenamefont {Bonitz},
  \citenamefont {Henning},\ and\ \citenamefont {Block}}]{9}%
  \BibitemOpen
  \bibfield  {author} {\bibinfo {author} {\bibfnamefont {M.}~\bibnamefont
  {Bonitz}}, \bibinfo {author} {\bibfnamefont {C.}~\bibnamefont {Henning}}, \
  and\ \bibinfo {author} {\bibfnamefont {D.}~\bibnamefont {Block}},\
  }\href@noop {} {\bibfield  {journal} {\bibinfo  {journal} {Rep.\ Prog.\
  Phys.}\ }\textbf {\bibinfo {volume} {73}},\ \bibinfo {pages} {066501}
  (\bibinfo {year} {2010})}\BibitemShut {NoStop}%
\bibitem [{\citenamefont {Morfill}\ \emph {et~al.}(2006)\citenamefont
  {Morfill}, \citenamefont {Konopka}, \citenamefont {Kretschmer}, \citenamefont
  {Rubin-Zuzic}, \citenamefont {Thomas}, \citenamefont {Zhdanov},\ and\
  \citenamefont {Tsytovich}}]{10}%
  \BibitemOpen
  \bibfield  {author} {\bibinfo {author} {\bibfnamefont {G.~E.}\ \bibnamefont
  {Morfill}}, \bibinfo {author} {\bibfnamefont {U.}~\bibnamefont {Konopka}},
  \bibinfo {author} {\bibfnamefont {M.}~\bibnamefont {Kretschmer}}, \bibinfo
  {author} {\bibfnamefont {M.}~\bibnamefont {Rubin-Zuzic}}, \bibinfo {author}
  {\bibfnamefont {H.~M.}\ \bibnamefont {Thomas}}, \bibinfo {author}
  {\bibfnamefont {S.~K.}\ \bibnamefont {Zhdanov}}, \ and\ \bibinfo {author}
  {\bibfnamefont {V.}~\bibnamefont {Tsytovich}},\ }\href@noop {} {\bibfield
  {journal} {\bibinfo  {journal} {New J.\ Phys.}\ }\textbf {\bibinfo {volume}
  {8}},\ \bibinfo {pages} {7} (\bibinfo {year} {2006})}\BibitemShut {NoStop}%
\bibitem [{\citenamefont {Caliebe}\ \emph {et~al.}(2011)\citenamefont
  {Caliebe}, \citenamefont {Arp},\ and\ \citenamefont {Piel}}]{11}%
  \BibitemOpen
  \bibfield  {author} {\bibinfo {author} {\bibfnamefont {D.}~\bibnamefont
  {Caliebe}}, \bibinfo {author} {\bibfnamefont {O.}~\bibnamefont {Arp}}, \ and\
  \bibinfo {author} {\bibfnamefont {A.}~\bibnamefont {Piel}},\ }\href@noop {}
  {\bibfield  {journal} {\bibinfo  {journal} {Phys.\ of Plasmas}\ }\textbf
  {\bibinfo {volume} {18}},\ \bibinfo {pages} {073702} (\bibinfo {year}
  {2011})}\BibitemShut {NoStop}%
\bibitem [{\citenamefont {Piel}\ \emph {et~al.}(2008)\citenamefont {Piel},
  \citenamefont {Arp}, \citenamefont {Klindworth},\ and\ \citenamefont
  {Melzer}}]{12}%
  \BibitemOpen
  \bibfield  {author} {\bibinfo {author} {\bibfnamefont {A.}~\bibnamefont
  {Piel}}, \bibinfo {author} {\bibfnamefont {O.}~\bibnamefont {Arp}}, \bibinfo
  {author} {\bibfnamefont {M.}~\bibnamefont {Klindworth}}, \ and\ \bibinfo
  {author} {\bibfnamefont {A.}~\bibnamefont {Melzer}},\ }\href@noop {}
  {\bibfield  {journal} {\bibinfo  {journal} {Phys.\ Rev.\ E}\ }\textbf
  {\bibinfo {volume} {77}},\ \bibinfo {pages} {026407} (\bibinfo {year}
  {2008})}\BibitemShut {NoStop}%
\bibitem [{\citenamefont {Menzel}\ \emph {et~al.}(2011)\citenamefont {Menzel},
  \citenamefont {Arp},\ and\ \citenamefont {Piel}}]{13}%
  \BibitemOpen
  \bibfield  {author} {\bibinfo {author} {\bibfnamefont {K.~O.}\ \bibnamefont
  {Menzel}}, \bibinfo {author} {\bibfnamefont {O.}~\bibnamefont {Arp}}, \ and\
  \bibinfo {author} {\bibfnamefont {A.}~\bibnamefont {Piel}},\ }\href@noop {}
  {\bibfield  {journal} {\bibinfo  {journal} {Phys.\ Rev.\ E}\ }\textbf
  {\bibinfo {volume} {83}},\ \bibinfo {pages} {016402} (\bibinfo {year}
  {2011})}\BibitemShut {NoStop}%
\bibitem [{\citenamefont {Arp}\ \emph {et~al.}(2011)\citenamefont {Arp},
  \citenamefont {Caliebe},\ and\ \citenamefont {Piel}}]{14}%
  \BibitemOpen
  \bibfield  {author} {\bibinfo {author} {\bibfnamefont {O.}~\bibnamefont
  {Arp}}, \bibinfo {author} {\bibfnamefont {D.}~\bibnamefont {Caliebe}}, \ and\
  \bibinfo {author} {\bibfnamefont {A.}~\bibnamefont {Piel}},\ }\href@noop {}
  {\bibfield  {journal} {\bibinfo  {journal} {Phys.\ Rev.\ E}\ }\textbf
  {\bibinfo {volume} {83}},\ \bibinfo {pages} {066404} (\bibinfo {year}
  {2011})}\BibitemShut {NoStop}%
\bibitem [{\citenamefont {Schwabe}\ \emph {et~al.}(2008)\citenamefont
  {Schwabe}, \citenamefont {Zhdanov}, \citenamefont {Thomas}, \citenamefont
  {Ivlev}, \citenamefont {Rubin-Zuzic}, \citenamefont {Morfill}, \citenamefont
  {Molotkov}, \citenamefont {Lipaev}, \citenamefont {Fortov},\ and\
  \citenamefont {Reiter}}]{15}%
  \BibitemOpen
  \bibfield  {author} {\bibinfo {author} {\bibfnamefont {M.}~\bibnamefont
  {Schwabe}}, \bibinfo {author} {\bibfnamefont {S.~K.}\ \bibnamefont
  {Zhdanov}}, \bibinfo {author} {\bibfnamefont {H.~M.}\ \bibnamefont {Thomas}},
  \bibinfo {author} {\bibfnamefont {A.~V.}\ \bibnamefont {Ivlev}}, \bibinfo
  {author} {\bibfnamefont {M.}~\bibnamefont {Rubin-Zuzic}}, \bibinfo {author}
  {\bibfnamefont {G.~E.}\ \bibnamefont {Morfill}}, \bibinfo {author}
  {\bibfnamefont {V.~I.}\ \bibnamefont {Molotkov}}, \bibinfo {author}
  {\bibfnamefont {A.~M.}\ \bibnamefont {Lipaev}}, \bibinfo {author}
  {\bibfnamefont {V.~E.}\ \bibnamefont {Fortov}}, \ and\ \bibinfo {author}
  {\bibfnamefont {T.}~\bibnamefont {Reiter}},\ }\href@noop {} {\bibfield
  {journal} {\bibinfo  {journal} {New J.\ Phys.}\ }\textbf {\bibinfo {volume}
  {10}},\ \bibinfo {pages} {033037} (\bibinfo {year} {2008})}\BibitemShut
  {NoStop}%
\bibitem [{\citenamefont {Morfill}\ \emph {et~al.}(1999)\citenamefont
  {Morfill}, \citenamefont {Thomas}, \citenamefont {Konopka}, \citenamefont
  {Rothermel}, \citenamefont {Zuzic}, \citenamefont {Ivlev},\ and\
  \citenamefont {Goree}}]{16}%
  \BibitemOpen
  \bibfield  {author} {\bibinfo {author} {\bibfnamefont {G.~E.}\ \bibnamefont
  {Morfill}}, \bibinfo {author} {\bibfnamefont {H.~M.}\ \bibnamefont {Thomas}},
  \bibinfo {author} {\bibfnamefont {U.}~\bibnamefont {Konopka}}, \bibinfo
  {author} {\bibfnamefont {H.}~\bibnamefont {Rothermel}}, \bibinfo {author}
  {\bibfnamefont {M.}~\bibnamefont {Zuzic}}, \bibinfo {author} {\bibfnamefont
  {A.}~\bibnamefont {Ivlev}}, \ and\ \bibinfo {author} {\bibfnamefont
  {J.}~\bibnamefont {Goree}},\ }\href@noop {} {\bibfield  {journal} {\bibinfo
  {journal} {Phys.\ Rev.\ Lett.}\ }\textbf {\bibinfo {volume} {83}},\ \bibinfo
  {pages} {1598} (\bibinfo {year} {1999})}\BibitemShut {NoStop}%
\bibitem [{\citenamefont {Khrapak}\ \emph {et~al.}(2011)\citenamefont
  {Khrapak}, \citenamefont {Klumov}, \citenamefont {Huber}, \citenamefont
  {Molotkov}, \citenamefont {Lipaev}, \citenamefont {Naumkin}, \citenamefont
  {Thomas}, \citenamefont {Ivlev}, \citenamefont {Morfill}, \citenamefont
  {Petrov}, \citenamefont {Fortov}, \citenamefont {Malentschenko},\ and\
  \citenamefont {Volkov}}]{17}%
  \BibitemOpen
  \bibfield  {author} {\bibinfo {author} {\bibfnamefont {S.~A.}\ \bibnamefont
  {Khrapak}}, \bibinfo {author} {\bibfnamefont {B.~A.}\ \bibnamefont {Klumov}},
  \bibinfo {author} {\bibfnamefont {P.}~\bibnamefont {Huber}}, \bibinfo
  {author} {\bibfnamefont {V.~I.}\ \bibnamefont {Molotkov}}, \bibinfo {author}
  {\bibfnamefont {A.~M.}\ \bibnamefont {Lipaev}}, \bibinfo {author}
  {\bibfnamefont {V.~N.}\ \bibnamefont {Naumkin}}, \bibinfo {author}
  {\bibfnamefont {H.~M.}\ \bibnamefont {Thomas}}, \bibinfo {author}
  {\bibfnamefont {A.~V.}\ \bibnamefont {Ivlev}}, \bibinfo {author}
  {\bibfnamefont {G.~E.}\ \bibnamefont {Morfill}}, \bibinfo {author}
  {\bibfnamefont {O.~F.}\ \bibnamefont {Petrov}}, \bibinfo {author}
  {\bibfnamefont {V.~E.}\ \bibnamefont {Fortov}}, \bibinfo {author}
  {\bibfnamefont {Y.}~\bibnamefont {Malentschenko}}, \ and\ \bibinfo {author}
  {\bibfnamefont {S.}~\bibnamefont {Volkov}},\ }\href@noop {} {\bibfield
  {journal} {\bibinfo  {journal} {Phys.\ Rev.\ Lett.}\ }\textbf {\bibinfo
  {volume} {106}},\ \bibinfo {pages} {205001} (\bibinfo {year}
  {2011})}\BibitemShut {NoStop}%
\bibitem [{\citenamefont {Thomas}\ \emph {et~al.}(2008)\citenamefont {Thomas},
  \citenamefont {Morfill}, \citenamefont {Fortov}, \citenamefont {Ivlev},
  \citenamefont {Molotkov}, \citenamefont {Lipaev}, \citenamefont {Hagl},
  \citenamefont {Rothermel}, \citenamefont {Khrapak}, \citenamefont
  {Suetterlin}, \citenamefont {Rubin-Zuzic}, \citenamefont {Petrov},
  \citenamefont {Tokarev},\ and\ \citenamefont {Krikalev}}]{18}%
  \BibitemOpen
  \bibfield  {author} {\bibinfo {author} {\bibfnamefont {H.~M.}\ \bibnamefont
  {Thomas}}, \bibinfo {author} {\bibfnamefont {G.~E.}\ \bibnamefont {Morfill}},
  \bibinfo {author} {\bibfnamefont {V.~E.}\ \bibnamefont {Fortov}}, \bibinfo
  {author} {\bibfnamefont {A.~V.}\ \bibnamefont {Ivlev}}, \bibinfo {author}
  {\bibfnamefont {V.~I.}\ \bibnamefont {Molotkov}}, \bibinfo {author}
  {\bibfnamefont {A.~M.}\ \bibnamefont {Lipaev}}, \bibinfo {author}
  {\bibfnamefont {T.}~\bibnamefont {Hagl}}, \bibinfo {author} {\bibfnamefont
  {H.}~\bibnamefont {Rothermel}}, \bibinfo {author} {\bibfnamefont {S.~A.}\
  \bibnamefont {Khrapak}}, \bibinfo {author} {\bibfnamefont {R.~K.}\
  \bibnamefont {Suetterlin}}, \bibinfo {author} {\bibfnamefont
  {M.}~\bibnamefont {Rubin-Zuzic}}, \bibinfo {author} {\bibfnamefont {O.~F.}\
  \bibnamefont {Petrov}}, \bibinfo {author} {\bibfnamefont {V.~I.}\
  \bibnamefont {Tokarev}}, \ and\ \bibinfo {author} {\bibfnamefont {S.~K.}\
  \bibnamefont {Krikalev}},\ }\href@noop {} {\bibfield  {journal} {\bibinfo
  {journal} {New J.\ Phys.}\ }\textbf {\bibinfo {volume} {10}},\ \bibinfo
  {pages} {033036} (\bibinfo {year} {2008})}\BibitemShut {NoStop}%
\bibitem [{\citenamefont {Schwabe}\ \emph {et~al.}(2011)\citenamefont
  {Schwabe}, \citenamefont {Jiang}, \citenamefont {Zhdanov}, \citenamefont
  {Hagl}, \citenamefont {Huber}, \citenamefont {Ivlev}, \citenamefont {Lipaev},
  \citenamefont {Molotkov}, \citenamefont {Naumkin}, \citenamefont
  {S{\"{u}}utterlin}, \citenamefont {Thomas}, \citenamefont {Fortov},
  \citenamefont {Morfill}, \citenamefont {Skvortsov},\ and\ \citenamefont
  {Volkov}}]{19}%
  \BibitemOpen
  \bibfield  {author} {\bibinfo {author} {\bibfnamefont {M.}~\bibnamefont
  {Schwabe}}, \bibinfo {author} {\bibfnamefont {K.}~\bibnamefont {Jiang}},
  \bibinfo {author} {\bibfnamefont {S.}~\bibnamefont {Zhdanov}}, \bibinfo
  {author} {\bibfnamefont {T.}~\bibnamefont {Hagl}}, \bibinfo {author}
  {\bibfnamefont {P.}~\bibnamefont {Huber}}, \bibinfo {author} {\bibfnamefont
  {A.~V.}\ \bibnamefont {Ivlev}}, \bibinfo {author} {\bibfnamefont {A.~M.}\
  \bibnamefont {Lipaev}}, \bibinfo {author} {\bibfnamefont {V.~I.}\
  \bibnamefont {Molotkov}}, \bibinfo {author} {\bibfnamefont {V.~N.}\
  \bibnamefont {Naumkin}}, \bibinfo {author} {\bibfnamefont {K.~R.}\
  \bibnamefont {S{\"{u}}utterlin}}, \bibinfo {author} {\bibfnamefont {H.~M.}\
  \bibnamefont {Thomas}}, \bibinfo {author} {\bibfnamefont {V.~E.}\
  \bibnamefont {Fortov}}, \bibinfo {author} {\bibfnamefont {G.~E.}\
  \bibnamefont {Morfill}}, \bibinfo {author} {\bibfnamefont {A.}~\bibnamefont
  {Skvortsov}}, \ and\ \bibinfo {author} {\bibfnamefont {S.}~\bibnamefont
  {Volkov}},\ }\href@noop {} {\bibfield  {journal} {\bibinfo  {journal} {EPL}\
  }\textbf {\bibinfo {volume} {96}},\ \bibinfo {pages} {55001} (\bibinfo {year}
  {2011})}\BibitemShut {NoStop}%
\bibitem [{\citenamefont {Chang}\ \emph {et~al.}(2011)\citenamefont {Chang},
  \citenamefont {Tseng},\ and\ \citenamefont {I}}]{20}%
  \BibitemOpen
  \bibfield  {author} {\bibinfo {author} {\bibfnamefont {M.-C.}\ \bibnamefont
  {Chang}}, \bibinfo {author} {\bibfnamefont {Y.-P.}\ \bibnamefont {Tseng}}, \
  and\ \bibinfo {author} {\bibfnamefont {L.}~\bibnamefont {I}},\ }\href@noop {}
  {\bibfield  {journal} {\bibinfo  {journal} {Phys.\ of Plasmas}\ }\textbf
  {\bibinfo {volume} {18}},\ \bibinfo {pages} {033704} (\bibinfo {year}
  {2011})}\BibitemShut {NoStop}%
\bibitem [{\citenamefont {Samsonov}\ \emph {et~al.}(2000)\citenamefont
  {Samsonov}, \citenamefont {Goree}, \citenamefont {Thomas},\ and\
  \citenamefont {Morfill}}]{21}%
  \BibitemOpen
  \bibfield  {author} {\bibinfo {author} {\bibfnamefont {D.}~\bibnamefont
  {Samsonov}}, \bibinfo {author} {\bibfnamefont {J.}~\bibnamefont {Goree}},
  \bibinfo {author} {\bibfnamefont {H.~M.}\ \bibnamefont {Thomas}}, \ and\
  \bibinfo {author} {\bibfnamefont {G.~E.}\ \bibnamefont {Morfill}},\
  }\href@noop {} {\bibfield  {journal} {\bibinfo  {journal} {Phys.\ Rev.\ E}\
  }\textbf {\bibinfo {volume} {61}},\ \bibinfo {pages} {5557} (\bibinfo {year}
  {2000})}\BibitemShut {NoStop}%
\bibitem [{\citenamefont {Zhukhovitskii}\ \emph {et~al.}(2012)\citenamefont
  {Zhukhovitskii}, \citenamefont {Fortov}, \citenamefont {Molotkov},
  \citenamefont {Lipaev}, \citenamefont {Naumkin}, \citenamefont {Thomas},
  \citenamefont {Ivlev}, \citenamefont {Schwabe},\ and\ \citenamefont
  {Morfill}}]{22}%
  \BibitemOpen
  \bibfield  {author} {\bibinfo {author} {\bibfnamefont {D.~I.}\ \bibnamefont
  {Zhukhovitskii}}, \bibinfo {author} {\bibfnamefont {V.~E.}\ \bibnamefont
  {Fortov}}, \bibinfo {author} {\bibfnamefont {V.~I.}\ \bibnamefont
  {Molotkov}}, \bibinfo {author} {\bibfnamefont {A.~M.}\ \bibnamefont
  {Lipaev}}, \bibinfo {author} {\bibfnamefont {V.~N.}\ \bibnamefont {Naumkin}},
  \bibinfo {author} {\bibfnamefont {H.~M.}\ \bibnamefont {Thomas}}, \bibinfo
  {author} {\bibfnamefont {A.~V.}\ \bibnamefont {Ivlev}}, \bibinfo {author}
  {\bibfnamefont {M.}~\bibnamefont {Schwabe}}, \ and\ \bibinfo {author}
  {\bibfnamefont {G.~E.}\ \bibnamefont {Morfill}},\ }\href@noop {} {\bibfield
  {journal} {\bibinfo  {journal} {Phys.\ Rev.\ E}\ }\textbf {\bibinfo {volume}
  {86}},\ \bibinfo {pages} {016401} (\bibinfo {year} {2012})}\BibitemShut
  {NoStop}%
\bibitem [{\citenamefont {Ivlev}\ and\ \citenamefont
  {Zhukhovitskii}(2012)}]{25}%
  \BibitemOpen
  \bibfield  {author} {\bibinfo {author} {\bibfnamefont {A.~V.}\ \bibnamefont
  {Ivlev}}\ and\ \bibinfo {author} {\bibfnamefont {D.~I.}\ \bibnamefont
  {Zhukhovitskii}},\ }\href@noop {} {\bibfield  {journal} {\bibinfo  {journal}
  {Phys.\ Plasmas}\ }\textbf {\bibinfo {volume} {19}},\ \bibinfo {pages}
  {093703} (\bibinfo {year} {2012})}\BibitemShut {NoStop}%
\bibitem [{\citenamefont {Landau}\ and\ \citenamefont {Lifshitz}(1959)}]{24}%
  \BibitemOpen
  \bibfield  {author} {\bibinfo {author} {\bibfnamefont {L.~D.}\ \bibnamefont
  {Landau}}\ and\ \bibinfo {author} {\bibfnamefont {E.~M.}\ \bibnamefont
  {Lifshitz}},\ }\href@noop {} {\emph {\bibinfo {title} {Fluid Mechanics}}}\
  (\bibinfo  {publisher} {Pergamon Press},\ \bibinfo {address} {New York},\
  \bibinfo {year} {1959})\BibitemShut {NoStop}%
\bibitem [{\citenamefont {Goree}\ \emph {et~al.}(1999)\citenamefont {Goree},
  \citenamefont {Morfill}, \citenamefont {Tsytovich},\ and\ \citenamefont
  {Vladimirov}}]{26}%
  \BibitemOpen
  \bibfield  {author} {\bibinfo {author} {\bibfnamefont {J.}~\bibnamefont
  {Goree}}, \bibinfo {author} {\bibfnamefont {G.~E.}\ \bibnamefont {Morfill}},
  \bibinfo {author} {\bibfnamefont {V.~N.}\ \bibnamefont {Tsytovich}}, \ and\
  \bibinfo {author} {\bibfnamefont {S.~V.}\ \bibnamefont {Vladimirov}},\
  }\href@noop {} {\bibfield  {journal} {\bibinfo  {journal} {Phys.\ Rev.\ E}\
  }\textbf {\bibinfo {volume} {59}},\ \bibinfo {pages} {7055} (\bibinfo {year}
  {1999})}\BibitemShut {NoStop}%
\bibitem [{\citenamefont {S{\"{u}}tterlin}\ \emph {et~al.}(2009)\citenamefont
  {S{\"{u}}tterlin}, \citenamefont {Wysocki}, \citenamefont {Ivlev},
  \citenamefont {R{\"{a}}th}, \citenamefont {Thomas}, \citenamefont
  {Rubin-Zuzic}, \citenamefont {Goedheer}, \citenamefont {Fortov},
  \citenamefont {Lipaev}, \citenamefont {Molotkov}, \citenamefont {Petrov},
  \citenamefont {Morfill},\ and\ \citenamefont {L{\"{o}}wen}}]{27}%
  \BibitemOpen
  \bibfield  {author} {\bibinfo {author} {\bibfnamefont {K.~R.}\ \bibnamefont
  {S{\"{u}}tterlin}}, \bibinfo {author} {\bibfnamefont {A.}~\bibnamefont
  {Wysocki}}, \bibinfo {author} {\bibfnamefont {A.~V.}\ \bibnamefont {Ivlev}},
  \bibinfo {author} {\bibfnamefont {C.}~\bibnamefont {R{\"{a}}th}}, \bibinfo
  {author} {\bibfnamefont {H.~M.}\ \bibnamefont {Thomas}}, \bibinfo {author}
  {\bibfnamefont {M.}~\bibnamefont {Rubin-Zuzic}}, \bibinfo {author}
  {\bibfnamefont {W.~J.}\ \bibnamefont {Goedheer}}, \bibinfo {author}
  {\bibfnamefont {V.~E.}\ \bibnamefont {Fortov}}, \bibinfo {author}
  {\bibfnamefont {A.~M.}\ \bibnamefont {Lipaev}}, \bibinfo {author}
  {\bibfnamefont {V.~I.}\ \bibnamefont {Molotkov}}, \bibinfo {author}
  {\bibfnamefont {O.~F.}\ \bibnamefont {Petrov}}, \bibinfo {author}
  {\bibfnamefont {G.~E.}\ \bibnamefont {Morfill}}, \ and\ \bibinfo {author}
  {\bibfnamefont {H.}~\bibnamefont {L{\"{o}}wen}},\ }\href@noop {} {\bibfield
  {journal} {\bibinfo  {journal} {Phys.\ Rev.\ Lett.}\ }\textbf {\bibinfo
  {volume} {102}},\ \bibinfo {pages} {085003} (\bibinfo {year}
  {2009})}\BibitemShut {NoStop}%
\end{thebibliography}
\end{document}